\documentclass[12pt]{article}

\usepackage{graphicx}
\begin{document}

\begin{center}
{\large\bf Can a few fanatics influence the opinion of a large segment of a 
society?}

\bigskip

Dietrich Stauffer$^\dagger$ and Muhammad Sahimi$^*$

{\it Mork Family Department of Chemical Engineering and Materials Science, 
University of Southern California, Los Angeles, California 90089-1211, USA}

\end{center}

\bigskip

Models that provide insight into how extreme opinions about any social
phenomenon may spread in a society or at the global scale are of great current 
interest. A realistic model must account for the fact that globalization, 
internet, and other means of mass communications have given rise to scale-free 
(SF) networks of interactions between people. We carry out extensive 
simulations of a new model which takes into account the SF nature of the 
interactions network, and provides some key insights into the phenomenon. The 
insights include, (1) the existence of a fundamental difference between a 
hierarchical network whereby people are influenced by those that are higher in 
the hierarchy but not by those below them, and a symmetrical network where 
person-on-person influence works mutually, and (2) the key result that a few 
"fanatics" can influence a large fraction of the population either temporarily 
(in the hierarchical interaction networks) or permanently (in symmetrical 
interaction networks). Even if the fanatics themselves disappear, the 
population may still remain susceptible to the ideologies or opinion originally
advocated by them. The model is, however, general and applicable to any 
phenomenon for which there is a degree of enthusiasm, or susceptibility to, in 
the population.

\bigskip

\noindent PACS number(s): 02.50.Ey, 05.40.-a, 89.65.-s, 89.75.-k

\bigskip

\noindent\rule{11.3truecm}{0.01in}

\noindent $^\dagger$Present and permanent address: Institute for Theoretical 
Physics, Cologne University, D-50923 K\"oln, Germany.\\
\noindent $^*$corresponding author. Electronic address: moe$@$iran.usc.edu

\newpage

\begin{center}
{\bf I. INTRODUCTION}
\end{center}

Given the current political climate around the world, and the rise of extreme
ideologies in many parts of the globe, models that can provide insight into how
such ideologies and opinions spread in a society are clearly of great interest.
To develop such models, one should keep in mind two well-known facts:

(1) Globalization, the internet, and other modern means of long-distance 
communications (for example, fax and mobile phones) have given rise to 
scale-free (SF) networks of interactions between people [1]. In a SF network 
the probability distribution $f(k)$ for a node to have $k$ links to other nodes
follows a power law,
\begin{equation}
f(k) \sim k^{-\gamma}\;,
\end{equation}
where $\gamma$ is a parameter that describes the abundance of the hubs, i.e.,
nodes of the network with large degree of connectiveness. Many unusual 
properties of SF networks have been attributed to distribution (1). 

(2) Typically, although extreme ideologies are originally advocated by very 
small fringe groups or even just a few ``fanatics,'' experience over the past 
several decades indicates that such ideologies may continue to survive and even
thrive over time scales that may be very large.

It is, therefore, clearly important to understand the role of the interactions
network on the opinion of a population, and how it affects such antisocial 
behavior as terrorism. Moreover, it is equally important to understand the 
conditions under which extreme ideologies can thrive and survive for a long 
time. If such understandings can be developed, they may help in designing 
effective ways of confronting and addressing the problem of extreme ideologies.

In this paper we carry out computer simulations of a model in order to better 
understand the phenomenon of the spread of extreme ideologies in a society. The
model is used to study how the opinions of various segments of a population may
be influenced by the interactions among individuals, and how the connectivity 
of the interactions network influences the survival or disappearance of an 
opinion. In particular, we are interested in learning whether it is possible 
for a few fanatics to influence a large population and, if so, what factors 
control the phenomenon and may prolong its life time. To do so, we represent 
the network of interactions between people by a SF network [1] and study 
various scenarios that may affect the dynamics of the spreading of an opinion 
in a population.

The phenomenon that we study, and the model that we develop for it, belong, in 
principle, to a general class of problems that describe various epidemic 
processes. In particular, our model and work are motivated by the study of 
Castillo-Chavez and Song [2] (see below). Great efforts have been devoted for 
decades to understanding how certain epidemic diseases, such as HIV, spread 
throughout a society [3,4]. In particular, the so-called SIS 
(susceptible-infected-susceptible), SIR (susceptible-infected-removed), and
SEIR (susceptible-exposed-infected-recovered) models have been developed and
studied either in terms of differential equations that describe the rate of 
change of each group of the population, or in discrete forms on regular 
lattices, such as the square lattice. The long-term dynamics of these models, 
when studied in terms of differential equations (which represent a type of 
mean-field approximation) or on regular lattices, is relatively simple [5] and 
can be expressed in terms of two fixed points: Either the disease dies out, or 
a stable equilibrium is reached whereby the disease is endemic. A threshold 
condition determines which of the two fixed points is stable. More complex 
behavior may arise when, for example, the model contains a seasonal forcing. 
Generalizations to models in which the ill individuals have a continuum of 
states have also been made [6].

More recently, a few of such models have been examined in complex networks in 
order to understand some {\it social} phenomena. In particular, Zanette [7] 
examined the dynamics of an epidemiclike model for the spread of a ``rumor'' on
a small-world (SW) network. A SW network is constructed starting from a
one-dimensional lattice with periodic boundary conditions which, in effect, 
make the lattice a ring, where each node is connected to its $2k$ nearest
neighbors, i.e., to the $k$ nearest neighbors clockwise and counterclockwise
[8]. To introduce disorder into the network, each of the $k$ clockwise 
connections of each node $i$ is rewired with a probability $q$ to a 
randomly-selected node $j$ that does not belong to the ``neighborhood'' of $i$.
In this way, the lattice contains shortcuts between distant nodes. Zanette [7]
showed that his model exhibits a transition between regimes of localization and
propagation at a {\it finite} value of the network randomness $q$. Somewhat 
similiar work was carried out by Shao {\it et al.} [9] who studied how 
``blackmail'' propagates in a SW network. In contrast, Pastor-Satorras and 
Vespignani [10] showed that a dynamical model of spreading of epidemics does 
{\it not} exhibit any threshold behavior when studied in a SF network, {\it in
the limit of a network of infinite size}, hence demonstrating a crucial 
difference between spreading of an epidemic phenomenon in SW and SF networks, 
which is clerarly due to their completely different connectivity structures.

The plan of this paper is as follows. In the next Section we describe the
model. Section III contains the results and a discussion of their implications.

\begin{center}
{\bf II. THE MODEL}
\end{center}

In the model the entire population is divided into four fractions: The general 
population $G$, those portions of the population that are either susceptible 
to, or excited about, an opinion, which we denote, respectively, by $S$ and 
$E$, and the fanatics $F$ who always advocate an opinion. Initially, 
everyone belongs to $G$, except a core of fanatics which, unless otherwise 
specified, is assumed to be four (but can be generalized to any number), since 
the most interesting results are obtained with a few initial fanatics (see 
below). Then, people can change their opinions depending on the neighbours to 
whom they listen to. Members of the $S,\;E,$ and $F$ groups can convince people
in the $G$ group to change their opinion and become susceptible to the 
fanatics's opinion; members of the $E$ and $F$ groups can convince the $S$ 
group to become $E$; members of the $F$ group can convince the $E$ members to
convert to $F$, but members of the $S,\;E,$ and $F$ groups can also directly 
return to the general population $G$. The fanatics are created initially by 
some outside event which is not part of the model. All the opinion changes 
happen with a probability $p$ that can have any particular value if there is 
any evidence for it. Such a model can be applied not only to terrorism and 
other extreme opinions, but also to any other social phenomenon for which there
is a degree of enthusiasm, or susceptibility to, in a society. 

A model of opinion dynamics was proposed recently based on the percolation
model [11]. Another recent model [12] uses, similar to our work, SF networks, 
but its dynamics and the quantities that it studies are completely different 
from those of the model studied in this paper. The partition of the population 
and the probabilities of opinion change in our model are similar to the model 
of Castillo-Chavez and Song [2] who proposed a deterministic continnum model in
terms of a set of nonlinear differential equations, given by
\begin{displaymath}
\frac{dS(t)}{dt}=\beta_1CG-\frac{\beta_2S(E+F)}{C}-\gamma_1S\;,
\end{displaymath}
\begin{equation}
\frac{dE(t)}{dt}=\frac{\beta_2S(E+F)}{C}-\frac{\beta_3EF}{C}-\gamma_2E\;, 
\end{equation}
\begin{displaymath}
\frac{dF(t)}{dt}=\frac{\beta_3EF}{C}-\gamma_3 F\;,
\end{displaymath}
where the various coefficients, $\beta_i$ and $\gamma_i$, are constant, and 
$C=S+E+F=1-G$. Without loss of generality, one can set $\beta_1=1$ since, 
otherwise, it can be absorbed in the time scale. (Omitting the denominators
in the above model does not change the results.) For comparison, the dynamics 
of the SEIR model is described by [5]
\begin{displaymath}
\frac{dS(t)}{dt}=\mu G-(\mu+\lambda)S\;,
\end{displaymath}
\begin{equation}
\frac{dE(t)}{dt}=\lambda S-(\mu+\sigma)E\;,
\end{equation}
\begin{displaymath}
\frac{dF(t)}{dt}=\sigma E-(\mu+\nu)F\;,
\end{displaymath}
and, $\lambda=\beta F$, with the various parameters being constant. It is
clear that the dynamics of our model is, in the continuum limit, much more 
complex than that of the SEIR model, even though they both are nonlinear. 
Castillo-Chavez and Song [2] studied their continuum model in detail. 
Similarly, the SEIR model was studied by, for example, Lloyd and May [5].

The models expressed by the sets (2) and (3) compute average behavior over 
the entire population and do not deal with individuals. Such approximations
cannot answer, for example, the question of whether or how a few fanatics can 
convince an entire population about a certain opinion or proposition. They 
cannot also take into account the effect of the SF structure of the interaction
network between people. Discretizing the model using a regular lattice, such as
a square lattice, is also not realistic because the range of the interactions 
in such networks is limited. Instead, networks [1] between people or computers 
are described better as scale-free, and a network of the Barab\'asi-Albert (BA)
type is the most widespread. This is a complex network in which the probability
distribution for a node to have $k$ links to other nodes follows Eq. (1) with 
$\gamma=3$. In such networks, a few people (nodes or hubs) have many 
connections, most people have rather few, and there is no sharp boundary 
between these extremes. We note that power laws also hold for the probability 
of terror attacks [13]. 

In this paper we simulate and study the model that we described above in the BA
network which, to our knowledge, has never been done before on either the SW or
SF networks. The BA networks are built by starting with four nodes (people) all
connected to each other. Newcomers then join the network one after the other by
connecting to the already-existing four members, with a probability 
proportional to the number of connections the member already has. In our study 
we use two BA types of SF networks. One is the {\it hierarchical} network with 
directed connections [14,15], which is a history-dependent network in the sense
that a member only listens to and can be convinced by the four people who 
joined earlier and were selected by the member. The four people, who are higher
in the hierarchy than the new member, do not listen to the new network member 
(that is, they do not change their opinion as a result of talking to the new 
network members). This is presumably the way a group with a rigid hierarchical 
command structure operates. An example, in the political arena, is provided by
the communist parties in China and in the old Soviet Union. Thus, one has a 
hierarchy determined by who joins the group first. The second type of the 
network that we use is {\it symmetrical} in the sense that all the connected 
members may influence each other, which is the way a group with a flexible 
command structure and spread out throughout the globe may operate, so that even
if the top leaders (the original fanatics) are eliminated, the group and its 
influence on people's opinion may live on. We have already seen examples of 
such groups in the Middle East and Latin America.

To simulate our model on a SF network, and to do so in a way that corresponds
to continuum model of Castillo-Chavez and Song [2], we adopt the following 
rules:
\begin{displaymath}
\mbox{$G\to S$ with probability $\beta_1$, if the neighbour is $S,\;E$, 
or $F$}\;,
\end{displaymath}
\begin{displaymath}
\mbox{$F\to G$ with probability $\gamma_3$}\;,
\end{displaymath}
\begin{displaymath}
\mbox{$E\to G$ with probability $\gamma_2$, and $E\to F$ with probability 
$\beta_3/C$, if the selected neighbour is $F$}\;,
\end{displaymath}
\begin{displaymath}
\mbox{$S\to G$ with probability $\gamma_1$, and $S\to E$ with probability 
$\beta_2/C$, if the selected neighbour is $E$ or $F$}\;.
\end{displaymath}
Thus, no person is convinced by an empty neighbour to change opinion. In this 
paper we simulate the behavior of two different systems. In one, we assume
that, $\beta_i=\gamma_i=p$ ($i=1,\;2$, and 3), as one main goal of this paper 
is to study the effect of a few well-connected fanatics on the opinion of an 
entire population. In the second case, we allow $\beta_i\neq\gamma_i$, and 
study several cases that we believe may yield interesting results and insights 
into the behavior of the phenomenon. We use an SF network of the BA type and, 
thus, the parameter $\gamma$ in Eq. (1) takes on a fixed value of 3. The 
average connectivity $\langle k\rangle$ of the SF network that we use is, 
$\langle k\rangle=8$. We will not consider any other value of $\gamma$ in the 
present paper.

Since the behavior of the population depends on the individuals' opinion and 
not just on their sum over all the lattice sites, sequential updating was used 
to simulate the model in both types of the network. We start with four fanatics
on the network core while everybody else belongs to the general population $G$.
We assume that the initial four fanatics are charismatic leaders forming the 
initial core of the network and, thus, becoming well-connected later. We also
consider the cases in which the number of the initial fanatics is less than 
four (see below). Except when indicated otherwise, we use in all cases a
{\it single} realization of the system. The reason for doing so is that we
are not interested in the {\it average} behavior of {\it all} societies. Some 
societies are more susceptible to extreme opinions or ideologies than others, 
whereas averaging the results over many realizations (populations) might mask 
the results particular to a given population.

\begin{center}
{\bf III. RESULTS AND DISCUSSION}
\end{center}

Figure 1 shows the results using the hierarchical network. Here, we used the
probability $p=\beta_i=\gamma_i=1/2$. It indicates that in the first few time 
steps a few fanatics can convert more than a million people to being 
susceptible to their ideology in a population of 25 million, even though the 
number of the (converted) fanatics actually falls down in the first few steps. 
The $E$ and $F$ groups grow to much smaller percentages. Finally, the three 
groups, $S,\;E,$ and $F$ vanish, and everybody returns to the general 
population $G$. However, the $S$ and $E$ groups can survive much longer than 
the original fanatics; it is even possible that the fanatics die out 
accidentally after three time steps. Nevertheless, the avalanche that they set 
in motion stays on for a long time, which is in fact a well-known phenomenon 
for many extreme ideologies or groups that believe in them.

The survival of the $S$ and $E$ groups, instead of their eventual extinction 
in the hierarchical networks that Fig. 1 indicates, is possible in the
symmetric network. This is shown in Fig. 2. For a probability $\beta_i=\gamma_i
=p=1/2$ to return from the $S,\;E,$ and $F$ groups to the general population 
$G$, the fanatics decrease from 4 to 2 in the first time step and vanish 
afterwards; nobody becomes excited, but up to 100 people become susceptible for
some time, which is indicated by the continuous curve in Fig. 2. If, however, 
we reduce from 1/2 to 0.1 the probability $p$ of returning from the $S,\;E,$ 
and $F$ groups to $G$, then all the four populations (shown by symbols in Fig. 
2) survive as large fractions of the total population. If we further reduce the
probability $p$ to 0.01, we will obtain the same survival pattern. This is 
shown in Figure 3.

The question of survival of the susceptible people (spread of the opinion)
appears to depend on the value of $\gamma_i$ and on whether or not $\beta_i=
\gamma_i$. For example, Fig. 4 presents the results obtained with the symmetric
model in which $\beta_i=0.5$ and $\gamma_i=0.1$, indicating the same patterns 
as in Figs. 2 and 3. We also find that if we hold all the $\beta_i$ fixed, and 
vary $\gamma_i$, we obtain a type of transition in the behavior of the system 
in the following sense. As already shown, for low values of $\gamma_i$ (for 
example, $\gamma_i=0.1$) the susceptible people always survive, while for large
values (for example, $\gamma_i=0.5$) they always die out. We find that there is
a critical value $\gamma_c$ of $\gamma_i$ in the symmetric model {\it at} which
the susceptibility dies out sometimes (that is, in certain realizations of a 
population) but survives at other times (in other realizations). We have 
determined this critical value to be, $\gamma_c\simeq 0.43$. An example is 
presented in Fig. 5 which shows that a finite number of susceptible people
survive in one realization, while dies out in another, albeit in a complex and
seemingly oscillatory pattern.

The mutual reinforcement of opinions in symmetric networks, which is impossible
in the hierarchical networks, greatly increases their spread in a society. For 
comparison, Fig. 6 shows the results for the hierarchical network with the 
reduced probability $p=\beta_i=\gamma_i=0.1$ to return, indicating that even 
with such value of the probability $p$ everybody becomes normal (returns to
the general population) after some time, i.e., stops believing in the fanatics'
opinion.

For a {\it fixed} set of the parameters, the fate of the susceptible people 
(that is, survival as opposed to decay and eventual vanishing) is the same in 
every realization of the symmetric network. But, the pattern of the 
fluctuations in the number of such people, and the time scale over which it 
may vanish, might be quite different. The question, then, is whether one might 
have some type of universal data collapse for all values of the parameters. We 
investigated this issue by carrying out extensive simulations with the 
symmetric model, using several values of $p=\beta_i=\gamma_i$, and summing
the results over $10^3$ realizations of the network. Figure 7 presents the 
results where the time has been rescaled to $pt$. Scaling and data collapse
hold roughly for small values of $p$. This implies that for small $p$ a change 
in all the transition probabilities is merely a change in the time scale, which
appears to be plausible.

The great influence of the four initial fanatics stems from the fact that the
founders of the network (where the fanatics reside), numbers 1, 2, 3, and 4 in 
its history, are well connected. The later a person joins the interaction 
network (higher membership numbers), the smaller is, in general, the number of 
connections and, thus, the influence. This effect is demonstrated in Fig. 8 
where we show the results for the hierarchical model with 25 million people. 
The top curve shows how up to 5\% of the population becomes susceptible under 
the influence of numbers 1, 2, 3, and 4 (taking $p=\beta_i=\gamma_i=1/2$). If,
instead, network members 11, 12, 13, and 14 are taken as the original fanatics 
(which are not as well-connected as those in numbers 1, 2, 3, and 4), then less
than 1\% of the population becomes susceptible (second curve from above in 
Fig. 8). The lower curves show analogously how the influence of the initial 
four fanatics is reduced if we take them as the four that follow numbers $10^2,
\;10^3,\dots,10^7$ in the networks of 25 million people (nodes). 

Due to the nonlinearity of the model, the initial concentrations, $E(0)$,
$S(0)$, and $G(0)$, are important to its dynamics and, therefore, we have 
considered their effect. We studied the case in which everybody outside the 
initial core was initially, (a) susceptible ($S$); (b) excited ($E$); (c) 
fanatic ($F$), or (d) belonged to the general population ($G$), as before. The 
four core members were always the fanatics ($F$). We studied the model in the 
hierarchical SF network with 35 million nodes.

Figure 9 shows the results for the evolution of susceptible population $S$ in 
the four cases, with the probability $p=\beta_i=\gamma_i=1/2$. Except 
when the entire system (aside from the core four fanatics) is composed of 
susceptible people, the fraction of the $S$ population first increases, 
reaching a maximum, but then decreases essentially exponentially, even when 
everybody in the network is initially a fanatic. A similar phenomenon happens
to the excited population $E$, the results for which are shown in Fig. 10. Such
a behavior will not change if the probability $p$ is varied. For example, Fig. 
11 presents the results for the $E$ population obtained with $p=\beta_i=
\gamma_i=0.2$, while Fig. 12 shows those for the $S$ population with $p=0.8$. 
In all cases, the excited and susceptible populations eventually vanish. Even 
the population of the fanatics eventually vanishes in the hierarchical 
structure. For example, Fig. 13 shows the results for the fanatic population 
with $p=0.2$. The only effect that the probability of conversion $p$ has is the
time scale over which the populations of the excited, fanatic, or susceptible 
people eventually vanish. Therefore, in a hierarchical structure everybody will
eventually go back to the general population, and will neither be susceptible 
to nor excited about the opinion originally advocated by the core fanatics. The
most important aspect of these results is the robust nature of the model: 
Regardless of the initial composition of the network, the $E$, $F$, and $S$ 
segments of the population eventually die out, and everybody returns to the 
general population.

To see whether the connectivity and hierarchical structure of the network make 
any difference to the results shown in Figs. 9-13, we repeated the simulations
using the symmetrical SF network in which the influence of two connected nodes 
on each other is mutual. Figure 14 presents the results for the susceptible 
population with $p=\beta_i=\gamma_i=0.5$, which should be compared with those 
shown in Fig. 9. Similar to Fig. 9, the susceptible population in this case 
decreases over time, but the reduction, rather than being exponential as in 
Fig. 9, is rather complex and resembles a seemingly oscillatory pattern, which
is due to the feedback mechanism which is present in the symmetrical network.

All the results presented so far were obtained with four initial fanatics.
What happens if we have fewer initial fanatics? We carried out simulations
with the symmetric model with only one initial fanatic. Since there are four
sites in the network's core but only one initial fanatic, we repeated the
simulations four times using the {\it same} network, each time starting with
the fanatic in a different core site. We found that there can be two distinct
cases: In one case the entire population becomes normal after the first time
step, while in the other three cases one obtains the same general patterns as 
before. The results for the three cases are shown in Fig. 15 where we present 
the number of susceptible people using $p=\beta_i=\gamma_i=0.5$. They are
completely similar to the continuous curve shown in Fig. 2. Varying values
of the parameters does not change this pattern, namely, either the entire
population becomes normal after the first or first few steps, or one obtains
the same general patterns as those obtained with four initial fanatics.

How would the above results differ if we carried out the same simulations
but on the square lattice, which has a very limited interaction range and 
fixed (and low) connectivity?  We find that in an $L\times L$ square lattice
with four initial fanatics the extreme opinion does not spread at long times,
regardless of the values of $\beta_i$ and $\gamma_i$, which is in contrast with
what we find in the SF networks. However, if we start with an entire line of
size $L$ of fanatics, we recover the SF-type behavior, namely, the extreme
ideology may or may not survive at long times, depending on the values of the 
parameters $\beta_i$ and $\gamma_i$. Therefore, there is a fundamental 
difference between the spread of an extreme opinion or ideology in a network of
people with the SF structure, and one with the severely restricted topology of 
a square lattice and similar networks, hence demonstrating the significance
of the range of people-to-people interactions.

BA networks have a percolation threshold [10] vanishing as 1/log($N$) and thus
purely geometrically information can always spread through a large population.
But this is only a necessary and not a sufficient condition for opinion
spreading; as Fig.1 and the lower curve in Fig.2 indicate, opinions may also
die out instead of spreading.

\begin{center}
{\bf SUMMARY}
\end{center}

Although some previous works [16] had investigated the spreading of a state 
shared by a number of agents, none was in the context of the type of model that
we study in this paper, namely, a four-component interacting system with the 
interactions being via a SF network. In addition, we find important differences
between the influence of the hierarchical and symmetric networks on opinion 
dynamics. If the followers listen to the leaders but not the other way around
(hierarchical interaction network), then the ideas of the leaders will die out.
In the political arena a good example is provided by the communism as advocated
by the Soviet Union in which there was a rigid structure imposed by the
communist party and its top leadership.

If, on the other hand, the leaders also listen to their followers, then their
opinions may last long, even if the leaders themselves are eliminated. The 
closer the leaders are to the core of the network (the best connected part of 
the network), the higher is their impact on the general population. Examples,
in the political arena, are provided by extremist groups in the Middle East and
Latin America. This phenomenon is also similar to Ising magnets studied on SF 
networks [17], but different from other models of opinion dynamics [15] in the
sense that, the hierarchical network structure yields results that are very 
different from those obtained by the undirected, symmetric networks. 

We regard the possibility of a few people influencing a large fraction of the 
population, and the persistence of an opinion in a symmetrical SF network but 
not in a hierarchical one, as the main results of this paper. Further 
predictions of the model, a comparison with its continuum counterpart, and its 
simulation on regular two-dimensional lattices, is reported elsewhere 
[18].

\begin{center}
{\bf ACKNOWLEDGMENT}
\end{center}

We thank Shlomo Havlin for suggesting that we study the behavior of the system
by holding one set of the parameters (the $\beta_i$) fixed and varying the 
other one (the $\gamma_i$).

\newpage

\newcounter{bean}
\begin{list}%
{[\arabic{bean}]}{\usecounter{bean}\setlength{\rightmargin}{\leftmargin}}

\item R. Albert and A.L. Barab\'asi, Rev. Mod. Phys. {\bf 74}, 47 (2002);
J.F.F. Mendes and S.N. Dorogovtsev, {\it Evolution of Networks: From
Biological Nets to the Internet and the WWW} (Oxford University Press, London,
2003).

\item C. Castillo-Chavez and B. Song, in, {\it Bioterrorism - Mathematical 
Modeling Applications in Homeland Security}, edited by H.T. Banks and C. 
Castillo-Chavez (SIAM, Philadelphia, 2003), p. 155.

\item N.T.J. Bailey, {\it The Mathematical Theory of Infectious Disease}, 2nd
ed. (Griffin, London, 1975).

\item R.M. Anderson and R.M. May, {\it Infectious Diseases of Humans} (Oxford 
University Oxford, Oxford, 1991); J.D. Murray, {\it Mathematical Biology} 
(Springer, Berlin, 1993); V. Capasso, {\it Mathematical Structures of Epidemic 
Systems} (Springer, Berlin, 1993); {\it Epidemic Models}, edited by D. Mollison
(Cambridge University Press, Cambridge, 1995).

\item See, for example, I.B. Schwartz and H.L. Smith, J. Math. Biol. {\bf 18},
233 (1983); A.L. Lloyd and R.M. May, J. Theor. Biol. {\bf 179}, 1 (1996);
M.J. Keeling, P. Rohani, and B.T. Grenfell, Physica D {\bf 148}, 317 (2001);
M. Kamo and A. Sasaki, Physica D {\bf 165}, 228 (2002); for a review see,
H.W. Hethcote, SIAM Rev. {\bf 42}, 4999 (2000).

\item H.C. Tuckwell, L. Toubiana, and J.-F. Vibert, Phys. Rev. E {\bf 57},
2163 (1998); {\it ibid.} {\bf 61}, 5611 (2000); {\it ibid.} {\bf 64}, 041918
(2001).

\item D.H. Zanette, Phys. Rev. E {\bf 64}, 050901(R) (2001); {\it ibid.}
{\bf 65}, 041908 (2002). See also, D.J. Watts and S.H. Strogatz, Nature 
(London) {\bf 393}, 440 (1998);

\item D.J. Watts, {\it Small Worlds} (Princeton University Press, Princeton, 
1999).

\item Z.-G. Shao, J.-P. Sang, X.-W. Zou, Z.-J. Tan, and Z.-Z. Jin, Physica A
{\bf 351}, 662 (2005).

\item R. Pastor-Satorras and A. Vespignani, Phys. Rev. Lett. {\bf 86}, 3200
(2001); Phys. Rev. E {\bf 63}, 066117 (2001). See also, R. Cohen, K. Erez, D. 
ben-Avraham, and S. Havlin, Phys. Rev. Lett. {\bf 85}, 4626 (2000).

\item S. Galam, Eur. Phys. J. B {\bf 26}, 269 (2002); Physica A {\bf 330}, 139 
(2003); Phys. Rev. E {\bf 71}, 046123 (2005); S. Galam and A. Mauger, {\it 
ibid.} {\bf 323}, 695 (2003).

\item G. Weisbuch, G. Deffuant, and F. Amblard, Physica A {\bf 353}, 55 (2005).

\item A. Clauset and M. Young, Scale invariance in global terrorism, 
physics/0502014 at www.arXiv.org (2005).

\item M.E.J. Newman, S.H. Strogatz, and D.J. Watts, Phys Rev. E {\bf 64}, 
026118 (2001); S.N. Dorogovtsev, J.F.F. Mendes, and A.N. Sanukhin, {\it ibid.}
025101 (2001); A.D. S\'anchez, J.M. L\'opez and M.A. Rodr\'{\i}guez, Phys. Rev.
Lett. {\bf 88}, 048701 (2002). 

\item D. Stauffer and H. Meyer-Ortmanns, Int. J. Mod. Phys. C {\bf 15}, 241 
(2004).

\item B. Chopard, M. Droz, and S. Galam, Eur. Phys. J. B {\bf 16}, 575 (2000);
S. Galam and J.P. Radomski, Phys. Rev. E {\bf 63}, 51907 (2001).

\item M.A. Sumour and M.M. Shabat, Int. J. Mod. Phys. C {\bf 16}, 584 (2005).

\item D. Stauffer and M. Sahimi, Physica A {\bf 364}, 537 (2006).

\end{list}%

\newpage

\begin{figure}[hbt]
\begin{center}
\includegraphics[angle=-90,scale=0.6]{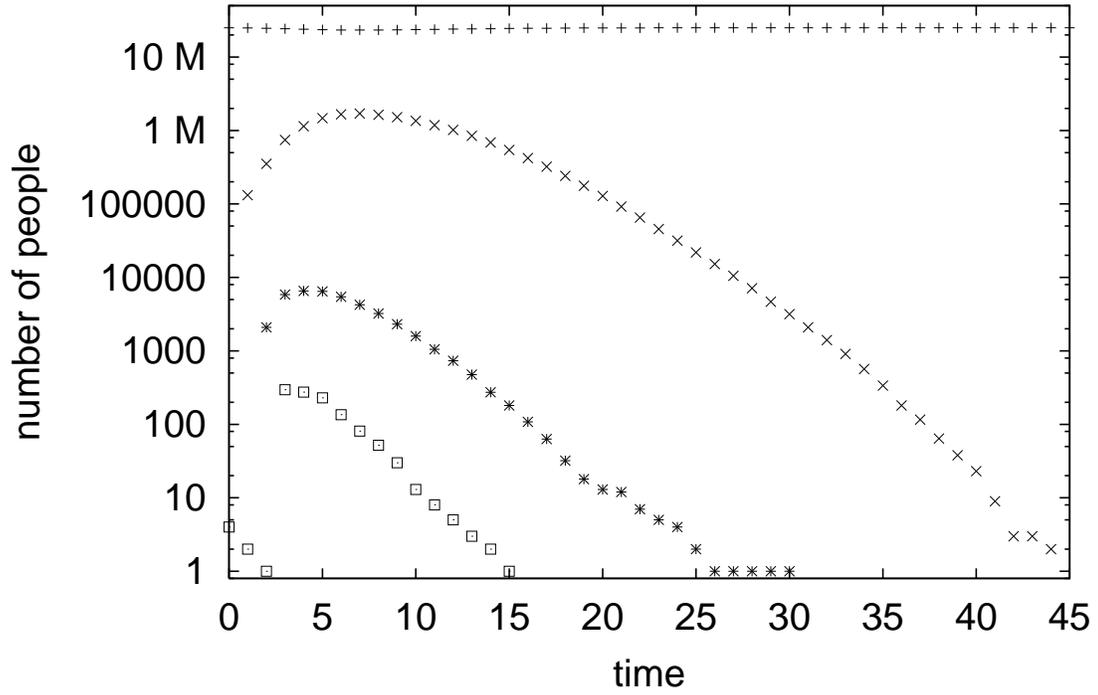}
\end{center}
\caption{Evolution, from top to bottom, of the general ($+$), the susceptible 
($\times$), the excited ($*$), and the fanatic population (squares) in the 
hierachical network. The total population is 25 million; the vertical scale is 
logarithmic, and $p=\beta_i=\gamma_i=0.5$.}
\end{figure}

\newpage

\begin{figure}[hbt]
\begin{center}
\includegraphics[angle=-90,scale=0.6]{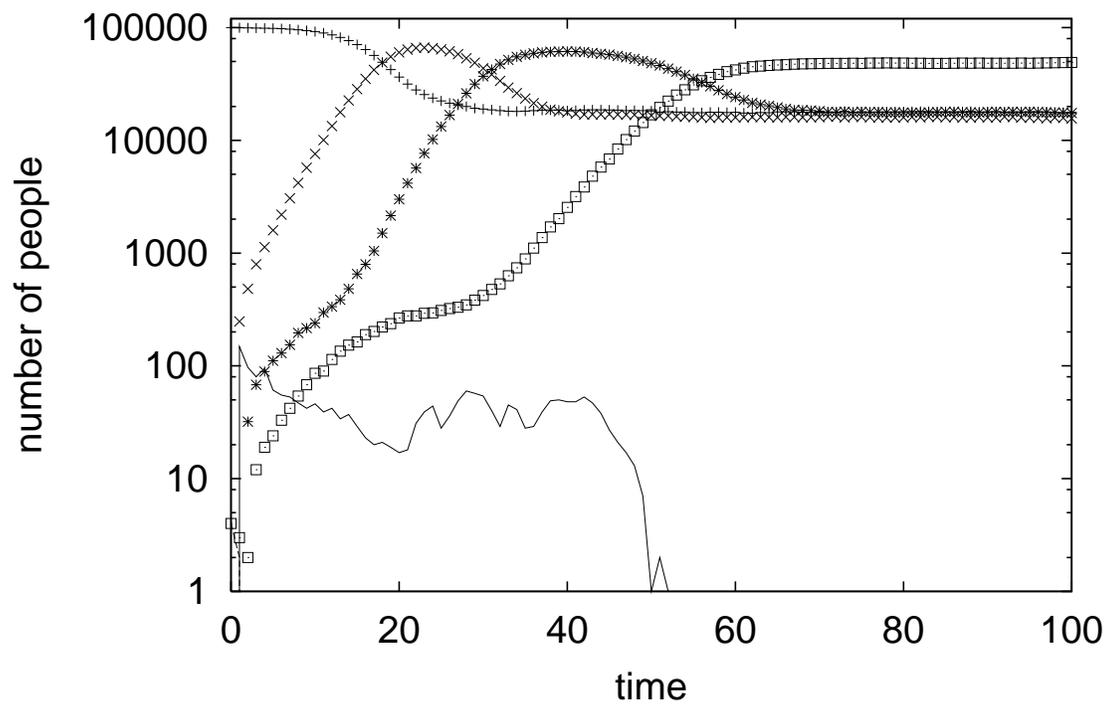}
\end{center}
\caption{
Symmetric network with $10^5$ people: With $p=\beta_i=\gamma_i=0.5$
the number of susceptible people (curve) first grows and then dies out. If, on
the other hand, the three $\gamma_i$ are reduced to 0.1, the $G$ (+), $S$
($\times$), $E$ ($*$) and $F$ (squares) groups all become roughly equal and do
{\it not} die out.
}
\end{figure}

\newpage

\begin{figure}[hbt]
\begin{center}
\includegraphics[angle=-90,scale=0.6]{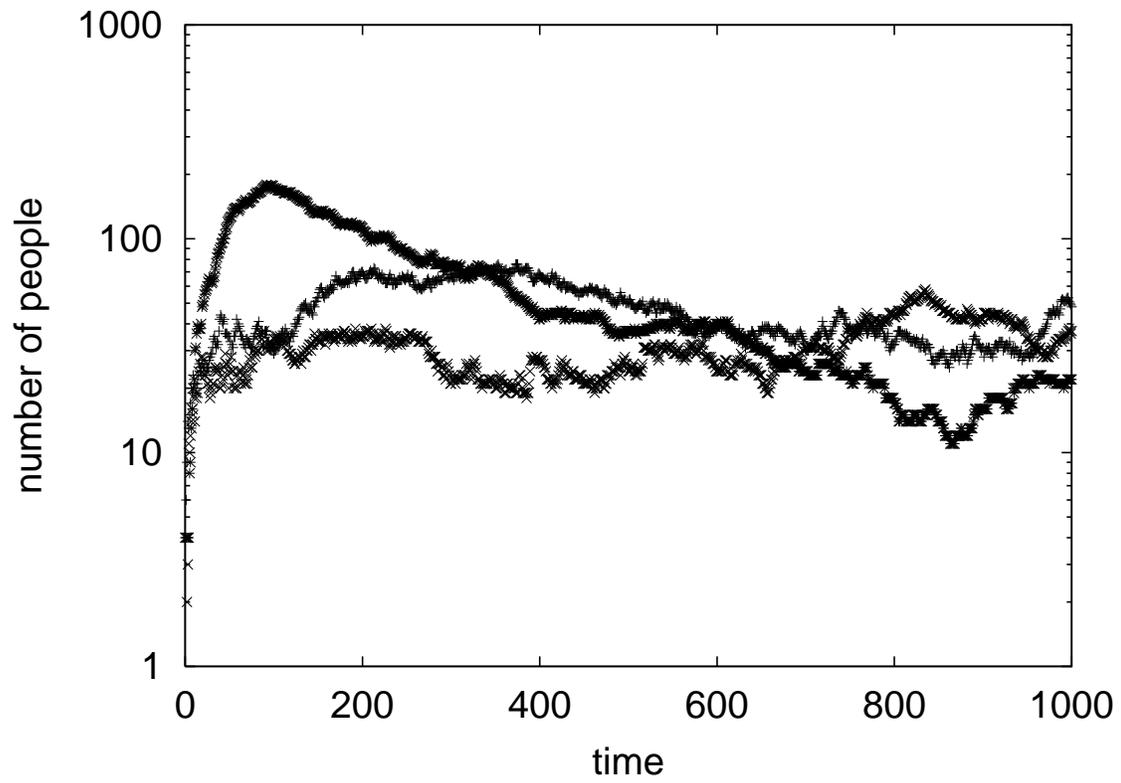}
\end{center}
\caption{Same as in Fig. 2, but with $p=\beta_i=\gamma_i=0.01.$}
\end{figure}

\newpage

\begin{figure}[hbt]
\begin{center}
\includegraphics[angle=-90,scale=0.6]{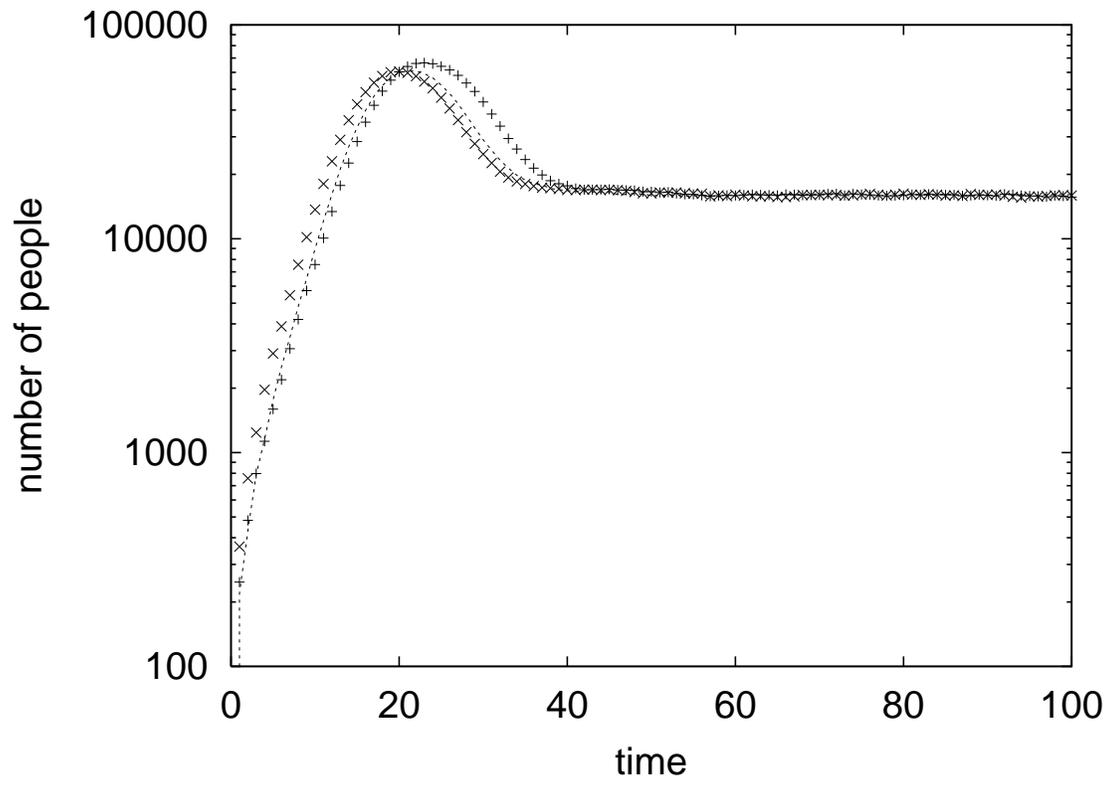}
\end{center}
\caption{Same as in Fig. 2, but with $\beta_i=0.5$ and $\gamma_i=0.1$.}
\end{figure}

\newpage

\begin{figure}[hbt]
\begin{center}
\includegraphics[angle=-90,scale=0.6]{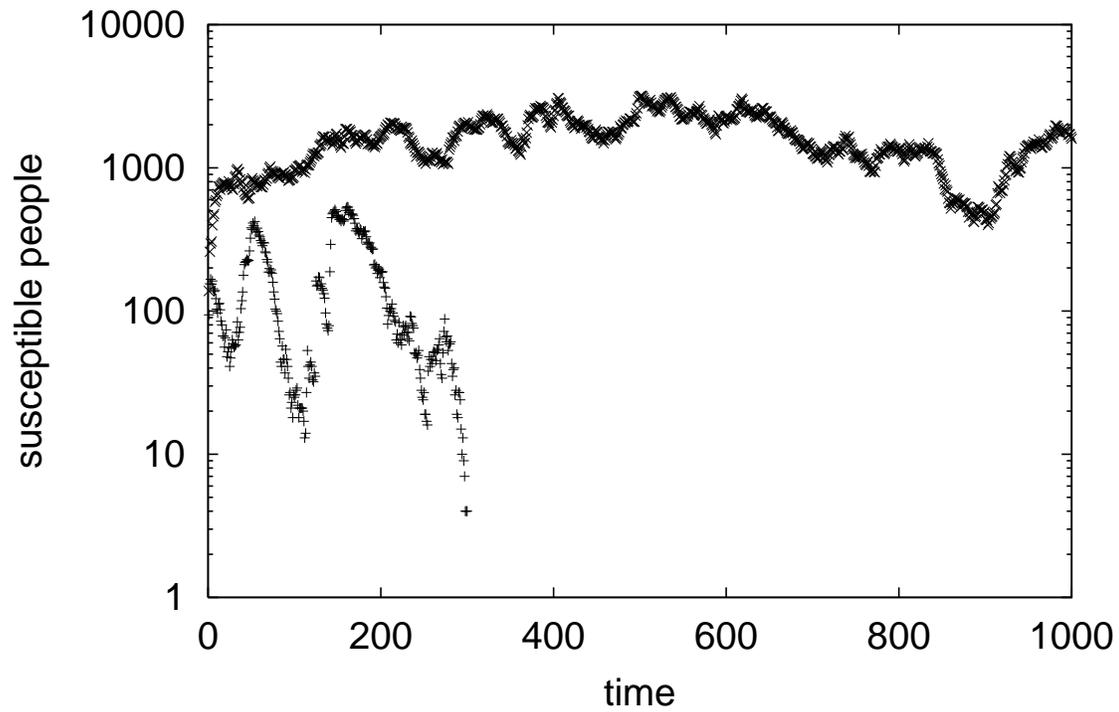}
\end{center}
\caption{Evolution of the number of susceptible people in the symmetric SF
network with $\beta_i=0.5$ and $\gamma_i=0.43$. In one realization (top) the
extreme ideology survives, while in the other (bottom) it eventually vanishes.}
\end{figure}

\newpage

\begin{figure}[hbt]
\begin{center}
\includegraphics[angle=-90,scale=0.6]{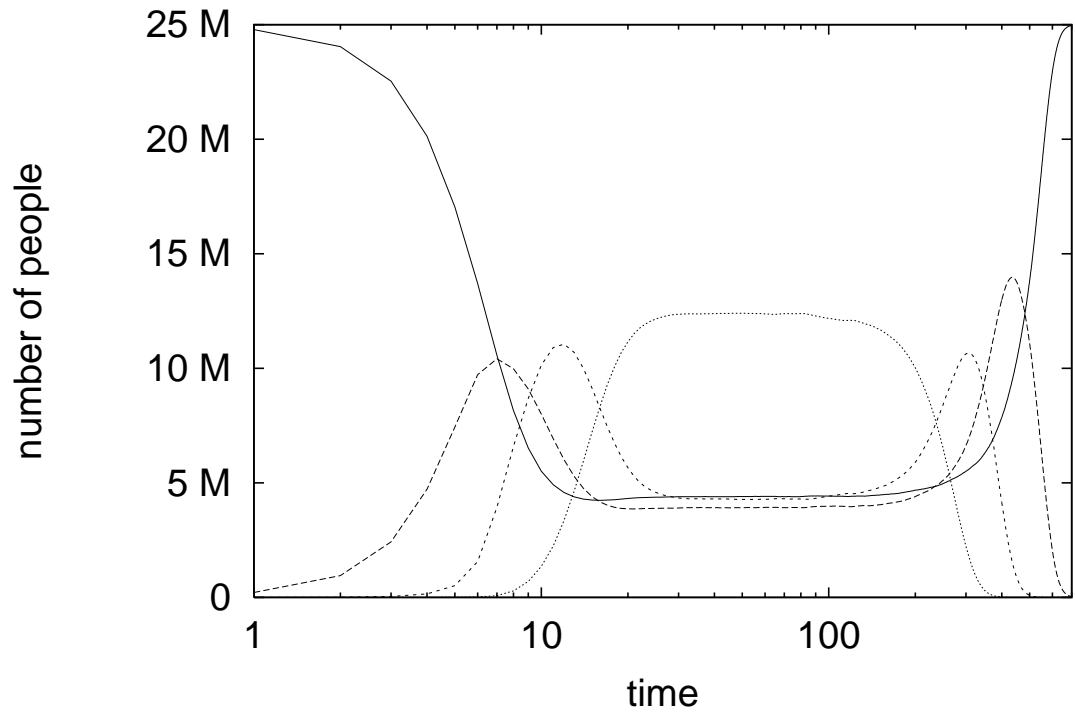}
\end{center}
\caption{The results for $\beta_i=\gamma_i=p=0.1$ for a hierarchical network of
25 million people, with linear vertical and logarithmic horizontal axes. 
Eventually, everyone returns to the general population $G$ as in Fig. 1. 
Symbols are the same as those in Fig. 2.}
\end{figure}

\newpage

\begin{figure}[hbt]
\begin{center}
\includegraphics[angle=-90,scale=0.6]{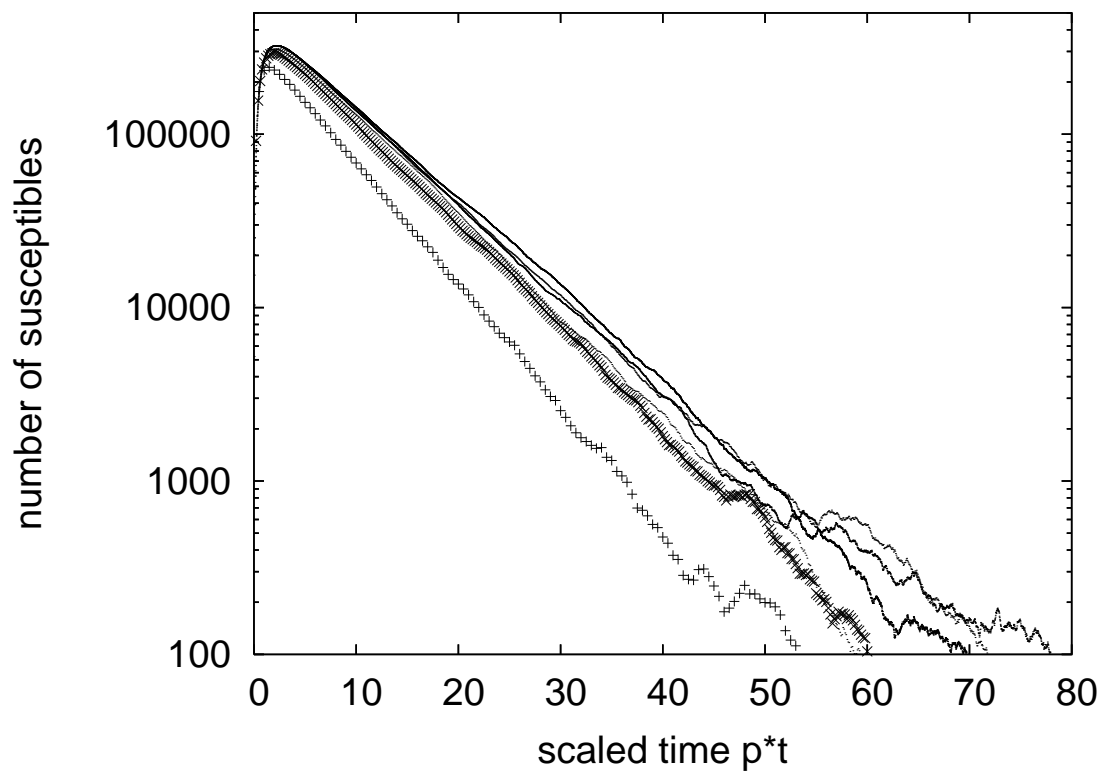}
\end{center}
\caption{Data collapse and scaling for the number of susceptible people versus 
the rescaled time $pt$, where $p=\beta_i=\gamma_i$. The results present the sum
over $10^3$ realizations of the symmetric networks of $10^5$ people for
$p=0.5\;(+),\; 0.2\;(\times),\;0.1,\;0.05,\;0.02,$ and 0.01. Data collapse 
holds roughly for small $p$.}
\end{figure}

\newpage

\begin{figure}[hbt]
\begin{center}
\includegraphics[angle=-90,scale=0.6]{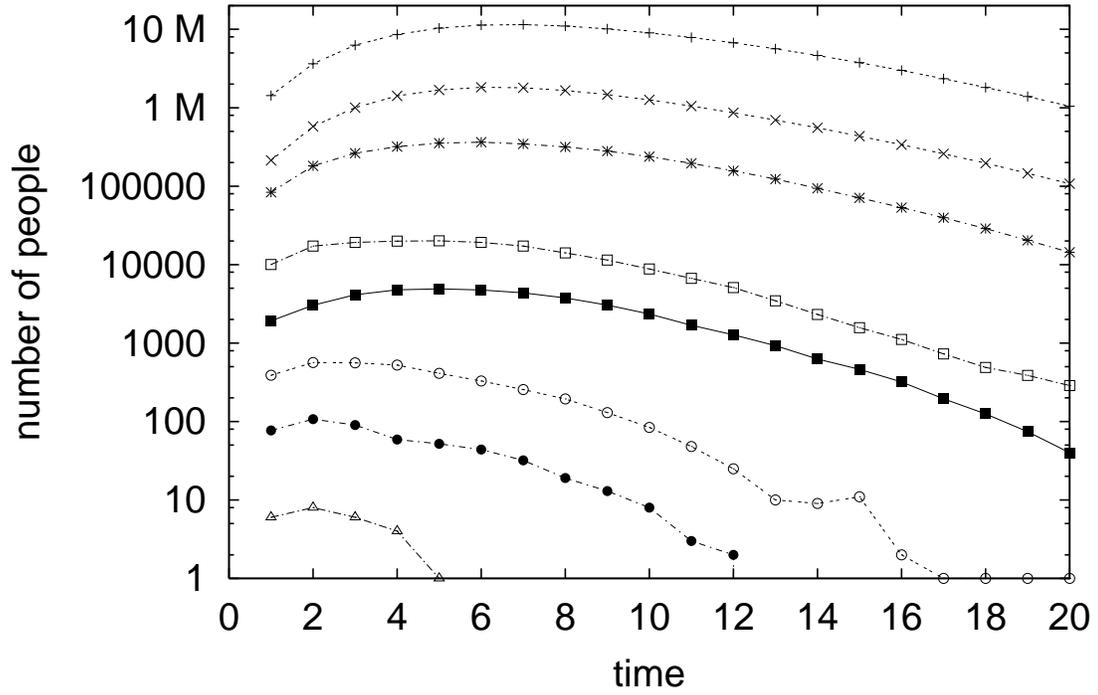}
\end{center}
\caption{The sum over 10 hierarchical networks of 25 million people each. 
The four initial radicals joined the network, from top to bottom, as numbers 1,
2, 3, and 4; 11, 12, 13, and 14; then 101, 102, 103, and 104, until 10000001, 
10000002, 10000003, 10000004 for the lowest curve. Latecomers are seen to have 
little influence. The results are for $p=\beta_i=\gamma_i=0.5$.}
\end{figure}

\newpage

\begin{figure}[hbt]
\begin{center}
\includegraphics[angle=-90,scale=0.6]{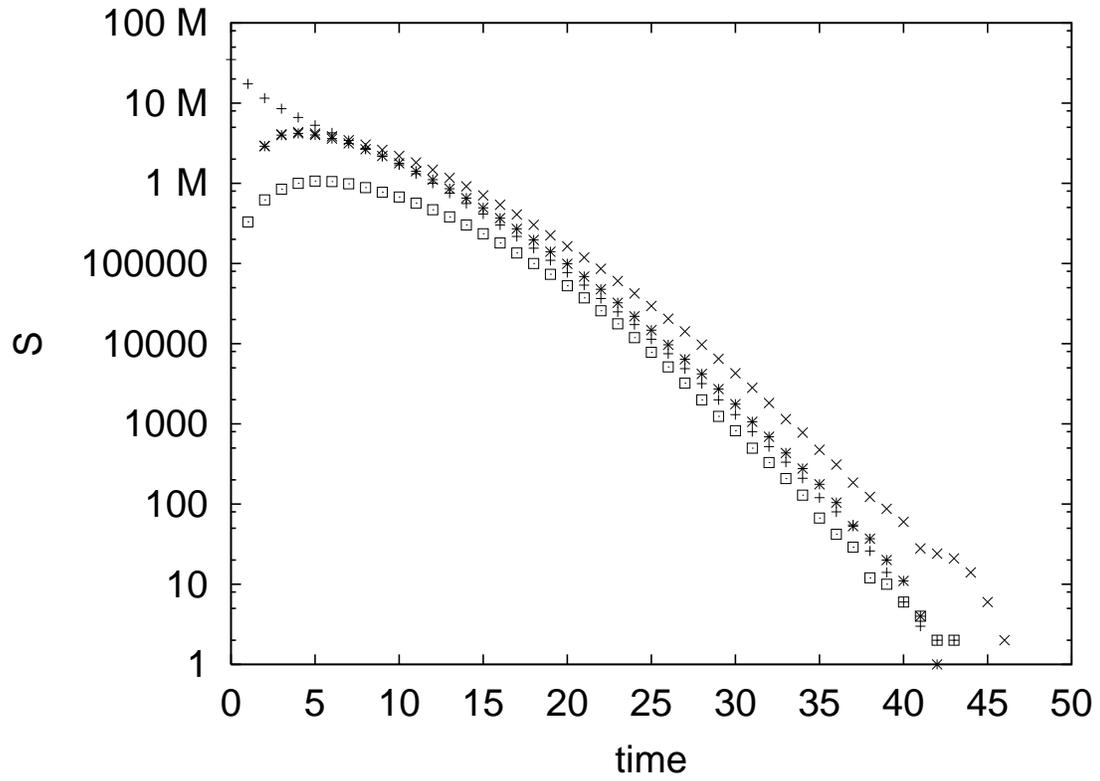}
\end{center}
\caption{The evolution of the susceptible population $S$ when, aside from the
core four fanatics, the rest of the population is initially susceptible ($+$),
exited ($\times$), fanatic ($*$), or belongs to the general population 
(squares). The results are for the hierarchical structure obtained with the 
probability $p=\beta_i=\gamma_i=1/2$.}
\end{figure}

\newpage

\begin{figure}[hbt]
\begin{center}
\includegraphics[angle=-90,scale=0.6]{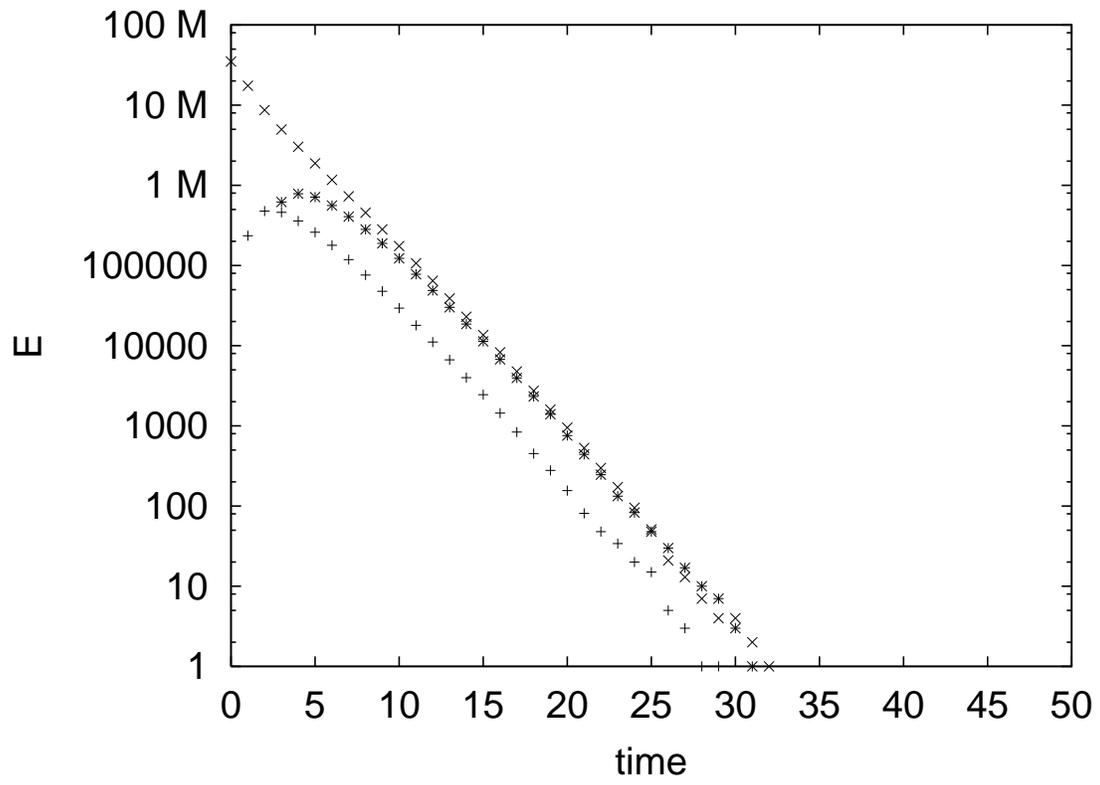}
\end{center}
\caption{Same as in FIG. 9, but for the excited population $E$.}
\end{figure}

\newpage

\begin{figure}[hbt]
\begin{center}
\includegraphics[angle=-90,scale=0.6]{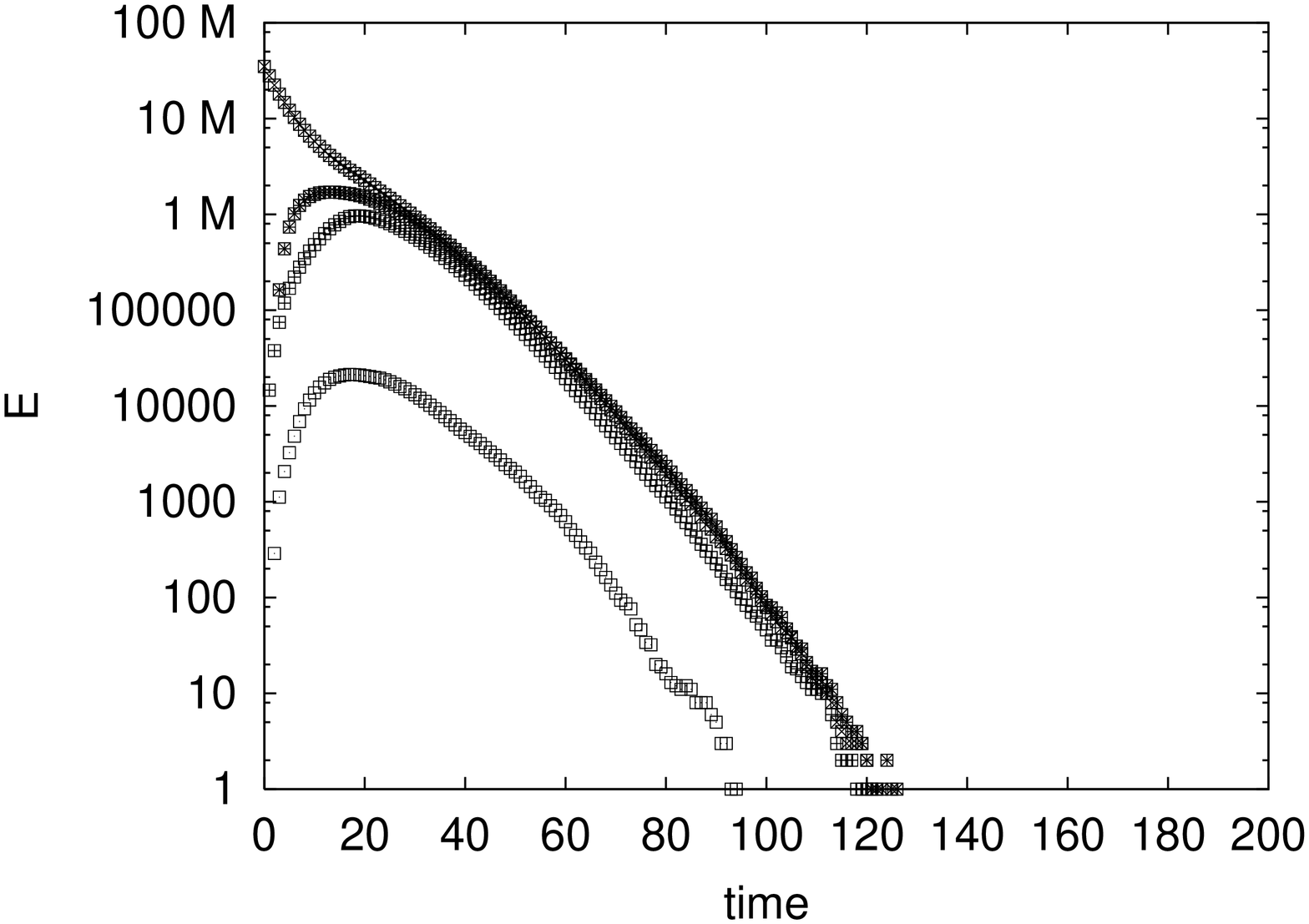}
\end{center}
\caption{Same as in FIG. 10, but with probability $p=\beta_i=\gamma_i=0.2$.}
\end{figure}

\newpage

\begin{figure}[hbt]
\begin{center}
\includegraphics[angle=-90,scale=0.6]{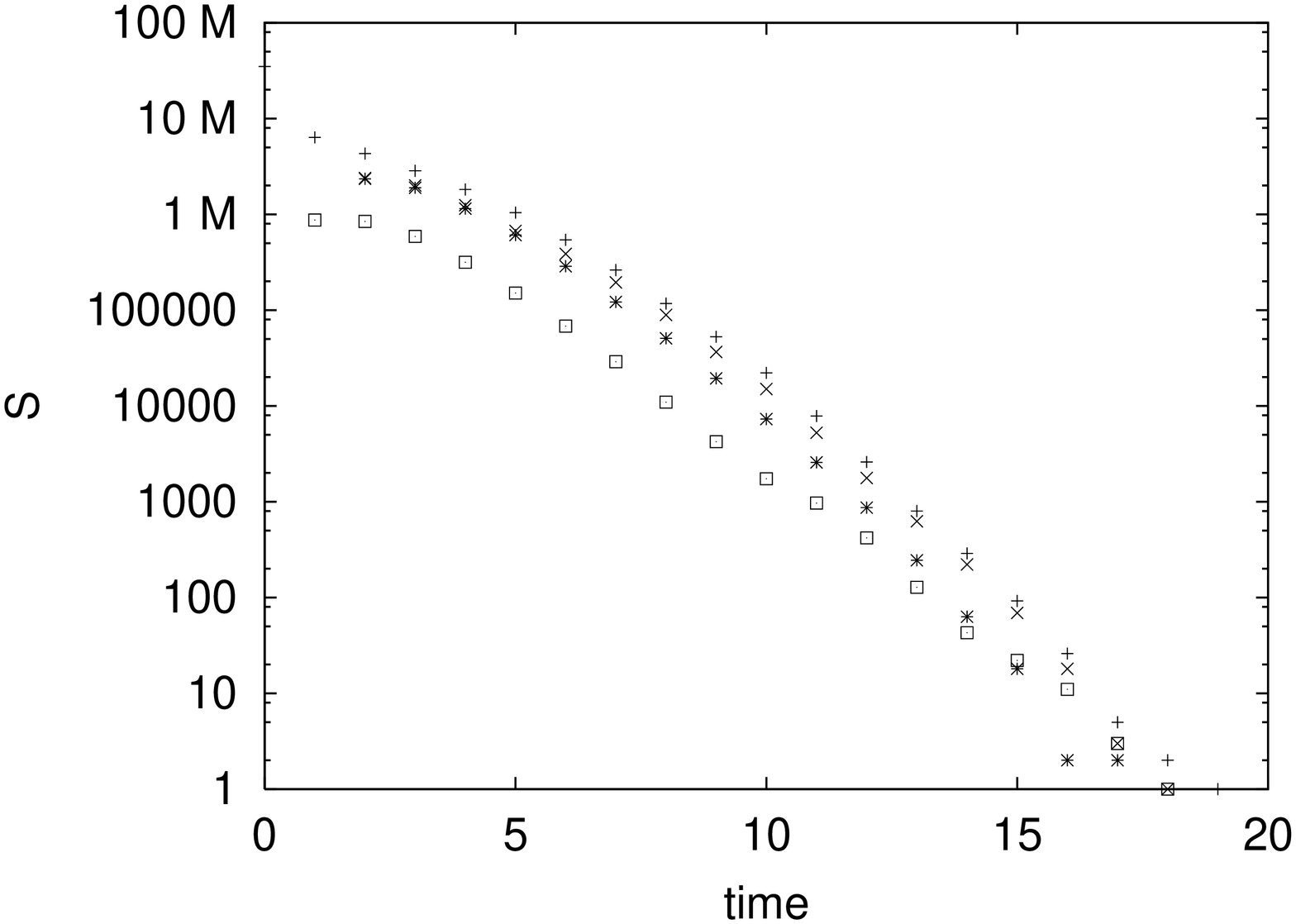}
\end{center}
\caption{Same as in FIG. 9, but with probability $p=\beta_i=\gamma_i=0.8$.}
\end{figure}

\newpage

\begin{figure}[hbt]
\begin{center}
\includegraphics[angle=-90,scale=0.6]{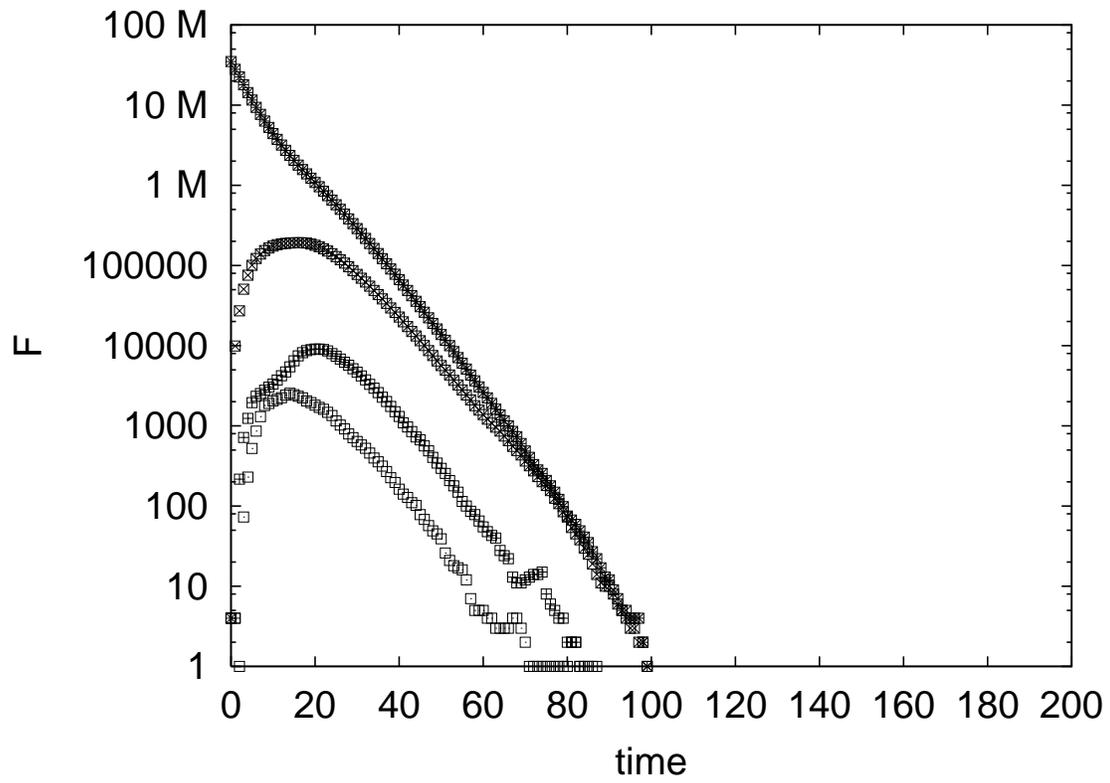}
\end{center}
\caption{Same as in FIG. 10, but for the fanatic population $F$.}
\end{figure}

\newpage

\begin{figure}[hbt]
\begin{center}
\includegraphics[angle=-90,scale=0.6]{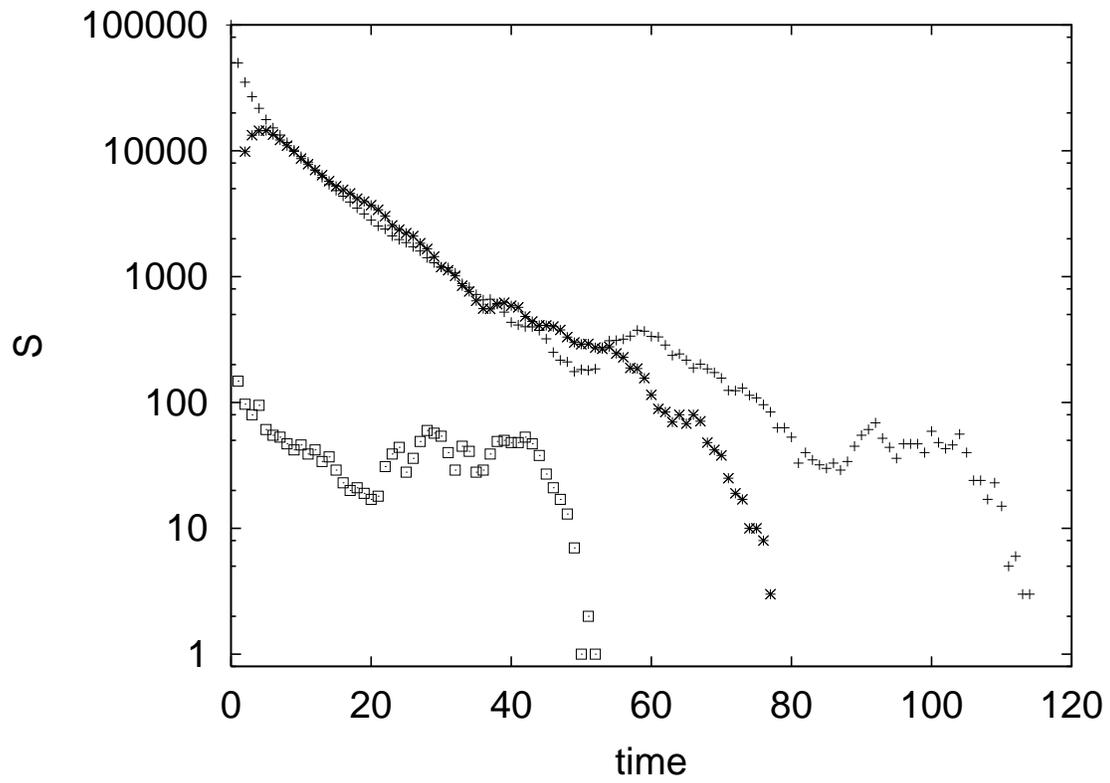}
\end{center}
\caption{Same as in FIG. 9, but obtained with the symmetrical SF network.}
\end{figure}

\newpage

\begin{figure}[hbt]
\begin{center}
\includegraphics[angle=-90,scale=0.6]{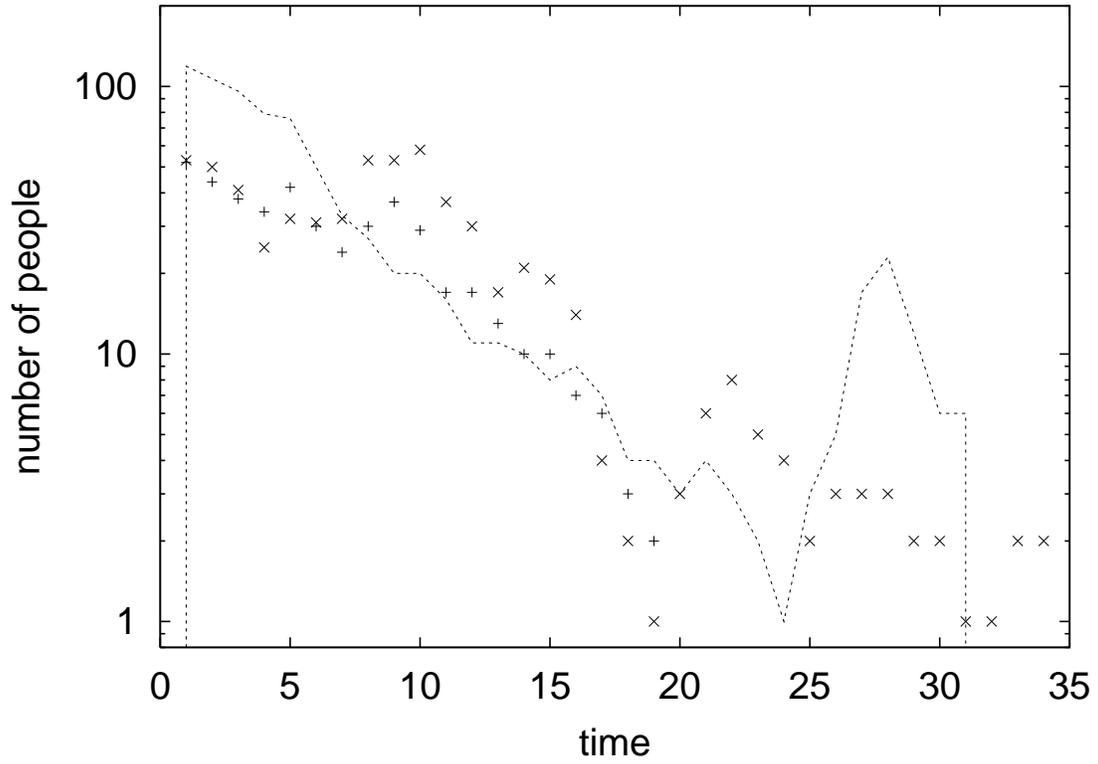}
\end{center}
\caption{Number of susceptible people obtained with the symmetrical
network with $p=\beta_i=\gamma_i=0.5$. Each of the symbols shows the results
for one of the four core sites of the network representing the initial single
fanatic. In the fourth case the entire population became normal after a single
time step and, hence, is not shown.}
\end{figure}

\end{document}